\documentclass[9pt]{article}
\usepackage{spconf}
\usepackage{textgreek}
\usepackage{cite}
\usepackage{xcolor}
\usepackage{upgreek}
\usepackage{epsfig}
\usepackage{svg}
\usepackage{float}
\usepackage{lipsum} 
\usepackage{multirow}
\usepackage{xcolor}
\usepackage{graphicx}
\usepackage{tabularx}
\usepackage{booktabs} 
\usepackage{float}
\usepackage{adjustbox}
\usepackage[hidelinks]{hyperref}

\title{HeightCeleb - an enrichment of VoxCeleb dataset with speaker height information}

\name{Stanisław Kacprzak and Konrad Kowalczyk\thanks{
 This work was supported in part by the National Science Centre, Poland, under Grant 2021/42/E/ST7/00452 and in part by the Excellence Initiative – Research University Program for the AGH University of Krakow. For the purpose of Open Access, the author has applied a CC-BY public copyright licence to any Author Accepted Manuscript (AAM) version arising from this submission.
 }}

\address{AGH University of Krakow, Institute of Electronics, 30-059 Krakow, Poland\\
\{skacprza, konrad.kowalczyk\}@agh.edu.pl}

\begin{document}

\maketitle

\begin{abstract}
Prediction of speaker's height is of interest for voice forensics, surveillance, and automatic speaker profiling. Until now, TIMIT has been the most popular dataset for training and evaluation of the height estimation methods. In this paper, we introduce HeightCeleb, an extension to VoxCeleb, which is the dataset commonly used in speaker recognition tasks. This enrichment consists in adding information about the height of all 1251 speakers from VoxCeleb that has been extracted with an automated method from publicly available sources. Such annotated data will enable the research community to utilize freely available speaker embedding extractors, pre-trained on VoxCeleb, to build more efficient speaker height estimators. In this work, we describe the creation of the HeightCeleb dataset and show that using it enables to achieve state-of-the-art results on the TIMIT test set by using simple statistical
regression methods and embeddings obtained with a popular speaker model (without any additional fine-tuning).
\end{abstract}
\setlength{\textfloatsep}{10pt}
\section{Introduction}
In the past two decades, there has been an increase of interest in extracting biometric information from speaker's voice. Most common application is speaker recognition, in which current state-of-the-art is to extract a speaker embedding using a deep neural network and compare its similarity with the embeddings gathered during enrollment \cite{Snyder2018, ecapa-tdnn}. Such systems, however, typically do not return any additional biometric cues.
On the other hand, speaker models could be applied to infer speaker's physical traits such as gender, height or age \cite{kalluri2020,ghahremani18b_interspeech,height_rajaa2021learning, height_gupta22_interspeech, age_voxceleb}. In this work, we focus on speaker's height, which has been shown to correlate with the length of the vocal tract e.g. in \cite{fitch1999morphology}. %
Information about the height of a speaker is relevant to voice forensics, surveillance, and automatic speaker profiling. It can help in criminal investigations, marketing campaigns or as auxiliary information in speaker identification in case of out-of-set individuals or unavailability of sufficient training data.

Early methods for height estimation consist in changing spectro-temporal features into various statistics such as mean, median or variances. For instance, features from Open-Smile \cite{eyben2010opensmile} toolkit are applied in \cite{mporas2009}, short-term cepstral features computed using a bag of words, at different time resolutions, are exploited in \cite{height_singh2016short}, while Mel Frequency Cepstral Coefficient (MFCC) features are used in \cite{kalluri2019}. Alternatively, fusion of spectral regression on analyzed formants and statistical acoustic model is investigated \cite{height_hansen2015speaker}. 
The appearance of deep learning has been the major game changer, and very quickly deep neural models started to be exploited to predict speaker's physical traits. A popular neural model for speaker embedding extraction known as Time Delay Neural Network (TDNN) \cite{Snyder2018} is applied for the height prediction task in \cite{data_nisp}, while fusion of the so-called x-vectors extracted from TDNN, with fixed-size embeddings known as i-vectors, are applied for a related task of age estimation in \cite{ghahremani18b_interspeech}. A significant gain in increasing accuracy of speaker profiling is achieved by approaches in which speaker representations are learned in a semi-supervised fashion, such as in \cite{height_rajaa2021learning} in which a CNN-LSTM encoder is applied for extracting speaker representation, or in a self-supervised fashion, such as in \cite{height_gupta22_interspeech} where the application of wav2vec 2.0 \cite{Baevski2020} model enables to capture more phonetic information. Note that the latter two approaches use semi-supervised or self-supervised learning as a remedy to the shortage of available labeled data. Such deep neural models are trained for the specific height estimation task or for multi-tasking purposes with a subset of height, age, and gender prediction tasks.

A major problem encountered when investigating methods to determine speaker's height from voice is the lack of datasets with height annotations. The most widely used, yet paid, dataset is TIMIT \cite{data_timit} with 630 English speakers. Alternatively, a free dataset with annotated height is NISP \cite{data_nisp}, with multilingual and multi accent data, but only 345 speakers. The scarcity of freely available annotated datasets is the key motivation for this research.

In this paper, we present HeightCeleb\footnote{\href{https://github.com/stachu86/HeightCeleb}{\url{https://github.com/stachu86/HeightCeleb}}} - an enrichment of the well-known and highly popular in speaker recognition VoxCeleb dataset \cite{data_voxceleb1}, in which we add information about speakers' height to the entire training and test sets. Our major goal is to provide the research community with significantly increased amount of annotated training data, and this way foster advancement of the research into height estimation from voice recordings. We describe how the HeightCeleb dataset was created and present its main height-related statistics such as mean, median, standard deviation, minimum and maximum values, as well as histograms for male and female speakers, and compare them with the existing datasets with height annotations. 

In addition, we show that this dataset enables to obtain competitive height prediction results using speaker embeddings pre-trained for the speaker recognition task (without any height-specific fine-tuning) and simple regression methods that are trained on HeightCeleb data. The results obtained for the TIMIT test set are comparable to those achieved by state-of-the-art methods reported in the literature. Finally, in order to evaluate the real usefulness of these methods, we would like to encourage to supplement the commonly used Mean Absolute Error and Root Mean Square Error metrics with additional statistical analysis, including at least the maximum error and probability of predicting height within a reasonable predefined range (e.g. 2\,cm).

\section{HeightCeleb dataset}
Motivated by the lack of freely available large speech datasets with information about speaker's height, we decided to annotate one ourselves. Since VoxCeleb \cite{data_voxceleb1} contains speech of celebrities, we suspected that gathering information about height of famous people should be relatively easy. A similar approach was taken in \cite{age_voxceleb} where authors enriched the VoxCeleb2 \cite{data_voxceleb2} datasets with age information. On a much smaller scale, to test their model, authors in \cite{height_hansen2015speaker} prepared a small dataset of 16 actors with speech and height from \emph{imdb.com}.

\subsection{Existing datasets with height information}
\begin{table}[t]
  \centering
  \caption{Statistics of datasets with height annotations in [cm]. %
  }
  \begin{adjustbox}{scale=0.95}
  \label{tab:height_statistics_all}
    \resizebox{\columnwidth}{!}{
    \begin{tabular}{ccccccccc}
    \toprule
    Dataset & Gender & Spk. No. & Mean & Median & Std & Min & Max \\
    \midrule
    \multirow{-2}{*}{} & Male & 438 & 179.68 & 180.0 & 7.0 & 157 & 203\\
    \multirow{-2}{*}{TIMIT} & Female & 192 & 165.83 & 165.0 & 6.8 & 145 & 183\\
    \midrule
    \multirow{-2}{*}{} & Male & 219 & 171.42 & 171.0 & 6.84 & 151 & 191\\
    \multirow{-2}{*}{NISP} & Female & 126 & 159.11 & 158.0 & 6.96 & 143 & 180\\
    \midrule
    \multirow{-2}{*}{} & Male & 690 & 180.32 & 180.0 & 7.04 & 157 & 208\\
    \multirow{-2}{*}{HeightCeleb} & Female & 561 & 166.49 & 166.0 & 6.99 & 145 & 192\\
    \bottomrule
  \end{tabular}}
  \end{adjustbox}
\end{table}
\begin{figure}[t]
  \centering
  \includeinkscape[width=1\columnwidth]{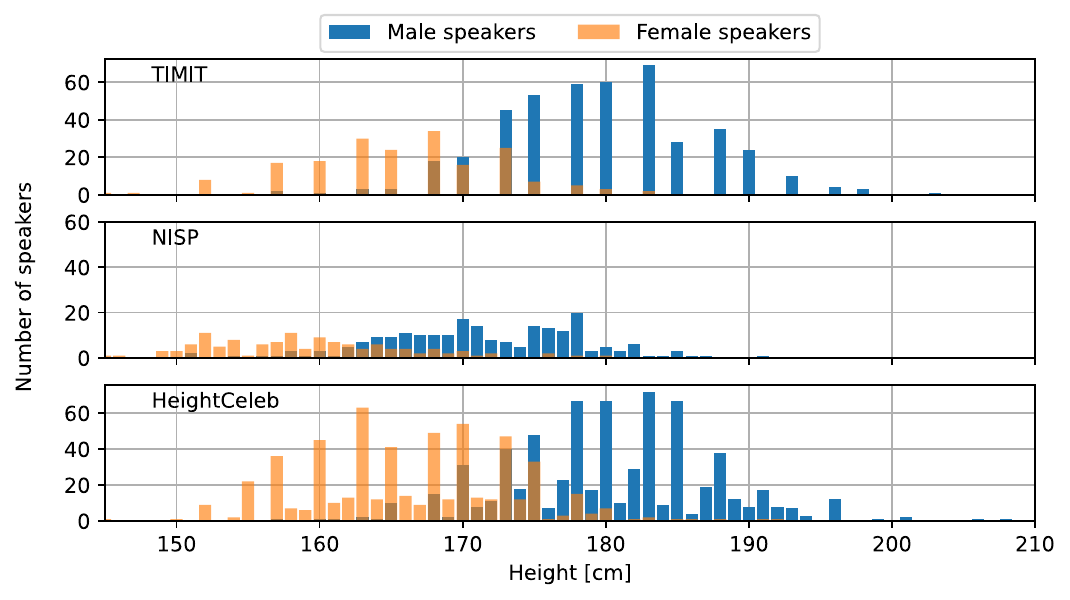}
  \caption{Speaker's height histograms across different datasets.}
  \label{fig:height_histogram_all}
\end{figure}
TIMIT is the most popular dataset with height annotations, which is composed of 2-5\,s long recordings of American English of 630 individuals. It contains 10 recordings per each of 438 male and 192 female speakers; it is thus strongly imbalanced with respect to gender. Most researchers use fixed train and test splits which have a similar proportion of male and female speakers. As noted in \cite{height_hansen2015speaker}, height information is self-reported, which has been shown in \cite{brestoff2011challenging} to likely cause an overestimation bias. Nonetheless authors of \cite{height_hansen2015speaker} sugest that TIMIT should not be strongly biased due to the lack of incentives to overestimate the height by anonymous individuals. Still, one needs to remember that besides intentional misreporting, there could be an error of measurement, temporal variation during the day, or reporting height measured many years ago \cite{brestoff2011challenging}. 
Speakers' height ranges from 145\,cm to 204\,cm, its distribution is shown in Figure \ref{fig:height_histogram_all}, while more detailed statistics in given in Table \ref{tab:height_statistics_all}. The comb-like spacing of histogram bars is the result of transforming the measurements performed with one inch precision, to the axis with one cm (only for the purpose of visualization). In all experiments, we use values without rounding.

Another dataset with annotated height is a freely available multi-language NISP dataset \cite{data_nisp}, which, however, contains only 345 speakers. NISP comprises speech in English and five native languages from India (Hindi, Kannada, Malayalam, Tamil and Telugu), with data on the weight, shoulder width and age. The height ranges from 143\,cm to 191\,cm, its distribution is also depicted in Figure \ref{fig:height_histogram_all}, while more detailed statistics is presented in Table \ref{tab:height_statistics_all}. A paid dataset with height annotations for a large number of speakers is NIST-SRE 2008 and 2010, which, as reported e.g. in \cite{Poorjam2015}, contains longer (even 10 minute long) recordings of 1236 speakers.

\subsection{Data collection of HeightCeleb}
\begin{figure}
\centering
\includegraphics[scale=0.2]{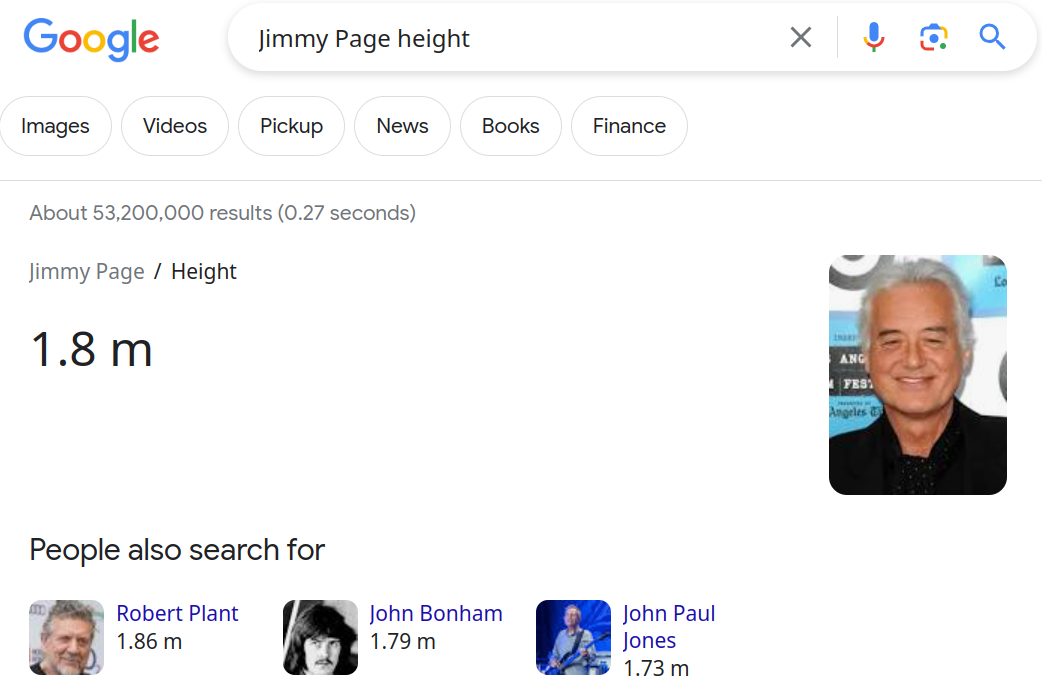}
\caption{Example of 'answer box' powered by Google~Knowledge~Graph, a prime source of the collected height data.}
\label{google_answer_box}
\end{figure}

To collect height information, we wrote a script that query \emph{google.com} search engine with the name of a person and the word 'height', and extracts height information from the \emph{answer box}, if returned in the results (see Figure~\ref{google_answer_box} for an example result). This information is provided from \emph{Google Knowledge Graph} and we treat it as our main source. %
As mentioned in the description of \emph{knowledge panels} (that are also powered by Knowledge Graph):\footnote{\href{https://support.google.com/knowledgepanel/answer/9163198}{\url{https://support.google.com/knowledgepanel/answer/9163198}}} \emph{'entities whose information is included in knowledge panels (like prominent individuals or the creators of a television show) are self-authoritative, and we provide ways for these entities to provide direct feedback'}. Therefore, some of the information displayed may also come directly from verified entities. Another adopted source of height information are websites \emph{imdb.com} and \emph{celebheights.com}, whereby the latter one is dedicated to the collection and estimation of height information of celebrities and contains \emph{'barefoot estimates, derived from quotations, official websites, agency resumes, in person encounters with actors at conventions and pictures/films.'}. %
Using those sources, we were able to collect almost all required data (1220 speakers using the main source and 26 speakers using two mentioned websites). We were determined to collect information for all speakers and allow the research community to simply use it in their experiment pipelines without any need for filtering or handling the missing values. Information about the 5 missing speakers was found by manually searching on the Internet. In case of differences between the sources, the subjective decision was made based on the quality of the source supported or by the analysis of pictures of a speaker.
\subsection{Reliability of the data collected for HeightCeleb}
It is extremely important to be aware of the limitations of collecting height information. The height data represents estimates, with difficult to determine errors. Beside obvious measurement and rounding errors (introduced when converting between imperial and metric systems), 
we suspect that height information may not always come from the same time when audio recording took place or it might be overestimated, as it is common that people tend to overestimate their height \cite{brestoff2011challenging}, and we also speculate that the speakers (celebrities) that appear in VoxCeleb may have greater insensitive for overestimation of height.
Note however, that this does not mean that the collected data is invaluable. On the contrary, we believe that any potential inaccuracies are still within a range that is sufficient for many practical applications, and the need for more precise height estimates would actually be very rare. Similar problems appear in TIMIT \cite{data_timit}, which also contains self-reported height information. The problems related to the reliability of the collected data are not unique to height information; for instance age annotation of Voxceleb2 in \cite{age_voxceleb} has estimated errors of up to 2 years, which is due to the lack of information about the precise dates of the recordings. %
To our mind, the collected HeightCeleb dataset should be used for training or validation, while tests should be performed on some \emph{gold standard} dataset with precise height measurements.
\subsection{Statistics of the height data in HeightCeleb} %
Since we enrich the data of the VoxCeleb dataset that contains the recordings of 1251 individuals of various length, the number of male and female speakers is the same as in the original VoxCeleb dataset. Compared with TIMIT, it is more balanced in terms of male-female parity, as it consists of roughly 55$\%$ male vs 45$\%$ female speakers. The distribution of height in the collected HeightCeleb dataset is depicted in Figure \ref{fig:height_histogram_all}, where it can be compared with an analogous distribution for the other two datasets. The height distribution of HeightCeleb is much denser than in TIMIT due to a nearly two times higher speaker number and values collected in cm, rather than collected in inches (and converted to cm) as done in case of TIMIT. The key statistical values for HeightCeleb are also presented in Table~\ref{tab:height_statistics_all}, and they can be compared with the numbers for the TIMIT and NISP datasets. It can be seen that the statistics (mean, median, min and max) of HeightCeleb and TIMIT are very similar.
Beside almost a doubled number of speakers, VoxCeleb contains between 45 to 250 utterances per speaker, with a total of over 153~thousands utterances recorded under real world conditions. Compared with NISP, HeightCeleb has a much higher number of speakers and a much better balanced height distribution, as it does not suffer from limitations brought about by the limited ethnic origins of speakers. Note that, in general, the mean and median for the HeightCeleb and TIMIT datesets are higher than for the NISP dataset by 6-9 cm.%
\section{Speaker's height prediction from speech}
\subsection{Proposed approach based on pre-trained neural speaker model and statistical regression}
\label{ssec:method}
We propose to predict speaker's height using a pre-trained neural speaker model followed a statistical regression method. In this work, speaker embedding is extracted using a popular neural speaker model called ECAPA-TDNN \cite{ecapa-tdnn}, which has been shown in the literature to be highly accurate in various speaker recognition tasks \cite{nagrani2020voxsrc,brown2022voxsrc,thienpondt2021}.
Contrary to the existing methods \cite{data_nisp,height_rajaa2021learning,height_gupta22_interspeech,ghahremani18b_interspeech}, which retrain the model using the regression based loss function, or a combination of the regression and other loss functions, we do not fine-tune the neural speaker model for the height estimation task. Instead, we rely on the pre-trained speaker model\footnote{\href{https://huggingface.co/speechbrain/spkrec-ecapa-voxceleb}{\url{https://huggingface.co/speechbrain/spkrec-ecapa-voxceleb}}} that was trained using VoxCeleb data, while we train only a simple statistical regression method to predict speaker's height directly from the extracted embedding. Our assumption is that a good speaker model should encode sufficient information about physical speaker traits such as height. Since there are many existing models pre-trained on VoxCeleb data, our HeightCeleb dataset will allow one to obtain reliable height estimates with embeddings obtained from those models. 

Since this work focuses on an introduction of a new dataset, we verify only two regression methods, namely Multiple Linear Regression (MLR) and Partial Least Squares Regression (PLSR) \cite{wold2001pls}, to show the potential of combining HeightCeleb dataset with the pre-trained speaker models. MLR does not require tuning of any hyper-parameters, while PLSR, which consists in projecting the predicted and observable variables to a new space of a lower dimensionality, takes the number of components (i.e. new dimensionality) as a hyper-parameter. Note that PLSR is suitable to data with multi-colinearity. 

To address the problem of the difference in height between genders, we follow an approach found in many existing works and train separate regression models for the male and female speakers, respectively. To this end, we apply a simple Logistic Regression based gender classifier. It is trained on embeddings estimated using ECAPA-TDNN from the TIMIT (train) set and achieves the accuracy of 99.2\% on the TIMIT (test) set, i.e. it is as accurate as a gender classification method presented in \cite{height_rajaa2021learning}.

\subsection{Performed experiments}
\begin{table*}[!ht]
\centering
\caption{Speaker height estimation results (MAE, RMSE, Max~Error) obtained using pre-trained ECAPA-TDNN embeddings and several investigated regression models for different combinations of training and test sets built upon TIMIT and HeightCeleb datasets.}
\begin{adjustbox}{scale=0.78}
\begin{tabular}{cccccc<{\hspace{3pt}}ccc<{\hspace{3pt}}ccc}
\toprule
\multicolumn{1}{c}{} & \multicolumn{1}{c}{} & \multicolumn{1}{c}{} & \multicolumn{2}{c}{} &\multicolumn{2}{c}{\textbf{MAE} {[}cm{]}} & \multicolumn{2}{c}{\textbf{RMSE} {[}cm{]}} & \multicolumn{2}{c}{\textbf{Max Error} {[}cm{]}}\\
\multicolumn{1}{c}{} & \multicolumn{1}{c}{\multirow{-2}{*}{Method}} & \multicolumn{1}{c}{\multirow{-2}{*}{Training set}} & \multicolumn{1}{c}{\multirow{-2}{*}{Test set}}  & \multicolumn{1}{c}{\multirow{-2}{*}{Test level}} & {Male} & {Female}  & {Male}    & {Female} & {Male} & {Female}\\ 
\midrule
\multicolumn{1}{c}{} & \multicolumn{1}{c}{\multirow{-2}{*}{}} & \multicolumn{1}{c}{\multirow{-2}{*}{}}& \multicolumn{1}{c}{\multirow{-2}{*}{}}& utterance & 5.34 & 5.24 & 7.01 & 6.52 & 23.36 & \textbf{14.98} \\
\multicolumn{1}{c}{} & \multicolumn{1}{c}{\multirow{-2}{*}{Baseline}} & \multicolumn{1}{c}{\multirow{-2}{*}{TIMIT (train)}} & \multicolumn{1}{c}{\multirow{-2}{*}{TIMIT (test)}} & speaker & 5.34 & 5.24 & 7.01 & 6.52 & 23.36 & 14.98 \\
\multicolumn{1}{c}{} & \multicolumn{1}{c}{\multirow{-2}{*}{}}& \multicolumn{1}{c}{\multirow{-2}{*}{}} & \multicolumn{1}{c}{\multirow{-2}{*}{}} & utterance & 5.14 & 5.52 & 6.55 & 6.95 & \textbf{20.24} & 20.11 \\ 
\multicolumn{1}{c}{} & \multicolumn{1}{c}{\multirow{-2}{*}{MLR}} & \multicolumn{1}{c}{\multirow{-2}{*}{TIMIT (train)}} & \multicolumn{1}{c}{\multirow{-2}{*}{TIMIT (test)}} & speaker & 4.69 & 5.11 & 6.04 & 6.41 & \textbf{15.93} & 16.28 \\
\multicolumn{1}{c}{} & \multicolumn{1}{c}{\multirow{-2}{*}{}}& \multicolumn{1}{c}{\multirow{-2}{*}{}} & \multicolumn{1}{c}{\multirow{-2}{*}{}} & utterance & 4.91 & 4.75 & 6.34 & \textbf{5.87} & 22.59 & 17.98 \\
\multicolumn{1}{c}{} & \multicolumn{1}{c}{\multirow{-2}{*}{MLR}} & \multicolumn{1}{c}{\multirow{-2}{*}{HeightCeleb (train)}} & \multicolumn{1}{c}{\multirow{-2}{*}{TIMIT (test)}} & speaker & 4.58 & 4.42 & 5.97 & \textbf{5.44} & 18.71 & \textbf{12.62} \\
\multicolumn{1}{c}{} & \multicolumn{1}{c}{\multirow{-2}{*}{}} & \multicolumn{1}{c}{\multirow{-2}{*}{}}& \multicolumn{1}{c}{\multirow{-2}{*}{}} & utterance & \textbf{4.77} & \textbf{4.56} & \textbf{6.16} & 5.89 & 22.50 & 16.59\\
\multicolumn{1}{c}{} & \multicolumn{1}{c}{\multirow{-2}{*}{PLSR}} & \multicolumn{1}{c}{\multirow{-2}{*}{HeightCeleb (train)}} & \multicolumn{1}{c}{\multirow{-2}{*}{TIMIT (test)}} & speaker & \textbf{4.53} & \textbf{4.40} & \textbf{5.87} & 5.72 & 18.11 & 14.28 \\
\midrule
\multicolumn{1}{c}{} & \multicolumn{1}{c}{\multirow{-2}{*}{}} & \multicolumn{1}{c}{\multirow{-2}{*}{}}& \multicolumn{1}{c}{\multirow{-2}{*}{}} & utterance  & 5.08 & 4.95 & 6.44 & 6.15 & 26.06 & 22.05 \\ 
\multicolumn{1}{c}{} & \multicolumn{1}{c}{\multirow{-2}{*}{MLR}} & \multicolumn{1}{c}{\multirow{-2}{*}{HeightCeleb (train)}}& \multicolumn{1}{c}{\multirow{-2}{*}{TIMIT (all)}} & speaker & 4.78 & 4.66 & 6.06 & 5.73 & 22.65 & 19.03 \\
\midrule
\multicolumn{1}{c}{} & \multicolumn{1}{c}{\multirow{-2}{*}{}} & \multicolumn{1}{c}{\multirow{-2}{*}{}} & \multicolumn{1}{c}{\multirow{-2}{*}{}} & utterance & 5.12 & 5.97 & 6.02 & 6.90 & 16.93 & 19.04 \\
\multicolumn{1}{c}{} & \multicolumn{1}{c}{\multirow{-2}{*}{MLR}} & \multicolumn{1}{c}{\multirow{-2}{*}{HeightCeleb (train)}}& \multicolumn{1}{c}{\multirow{-2}{*}{HeightCeleb (test)}} & speaker & 4.06 & 5.43 & 4.77 & \textbf{6.12} & 10.38 & 10.51 \\
\multicolumn{1}{c}{} & \multicolumn{1}{c}{\multirow{-2}{*}{}} & \multicolumn{1}{c}{\multirow{-2}{*}{}}& \multicolumn{1}{c}{\multirow{-2}{*}{}} & utterance & \textbf{4.59} & \textbf{5.44} & \textbf{5.42} & \textbf{6.39} & \textbf{15.42} & \textbf{15.93} \\ 
\multicolumn{1}{c}{} & \multicolumn{1}{c}{\multirow{-2}{*}{PLSR}} & \multicolumn{1}{c}{\multirow{-2}{*}{HeightCeleb (train)}}& \multicolumn{1}{c}{\multirow{-2}{*}{HeightCeleb (test)}} & speaker & \textbf{3.68} & \textbf{5.38} & \textbf{4.35} & \textbf{6.12} & \textbf{9.47} & \textbf{10.29} \\
\bottomrule
\end{tabular}
\end{adjustbox}
\label{tab:our_experiments}
\end{table*}

In all experiments, speaker embeddings (with dimensionality $d=192$) are obtained from the SpeechBrain implementation \cite{speechbrain} of the ECAPA-TDNN model \cite{ecapa-tdnn}, while implementations of both regression methods, namely MLR and PLSR, come from the scikit-learn library \cite{scikit-learn}). In addition, we show the results for the Baseline method which always predicts the mean of the training set for the respective gender (the method is equivalent to the regression method with the coefficient of determination $R^2=0$).
As evaluation measures, we present the commonly used Mean Absolute Error (MAE) and Root Mean Square Error (RMSE) metrics, as well as the Maximum Error observed for the entire test dataset. All measures are calculated independently for the male and female speakers. Furthermore, we report on two types of errors (denoted as test level), i.e. errors obtained individually per each utterance and errors averaged across all utterances of the same speaker. The per-speaker errors are rarely reported in the literature, however, we believe they are important from a practical point of view as some applications may use multiple samples to make the final decision.
For training of the regression method, we either use the TIMIT (train) set or the HeightCeleb (train) set. The height prediction tests are performed either on the TIMIT (test) set with 112 male and 56 female speakers (1120 and 560 utterances, respectively), on the HeightCeleb (test) set with 25 male and 15 female speakers (3242 and 1450 utterances, respectively), or on the TIMIT (all) set that includes both train and test splits, provided that training was performed on the HeightCeleb (train) set.
\subsection{Evaluation results and discussion}
\begin{figure}[t]
  \centering
  \includeinkscape[width=0.8\columnwidth]{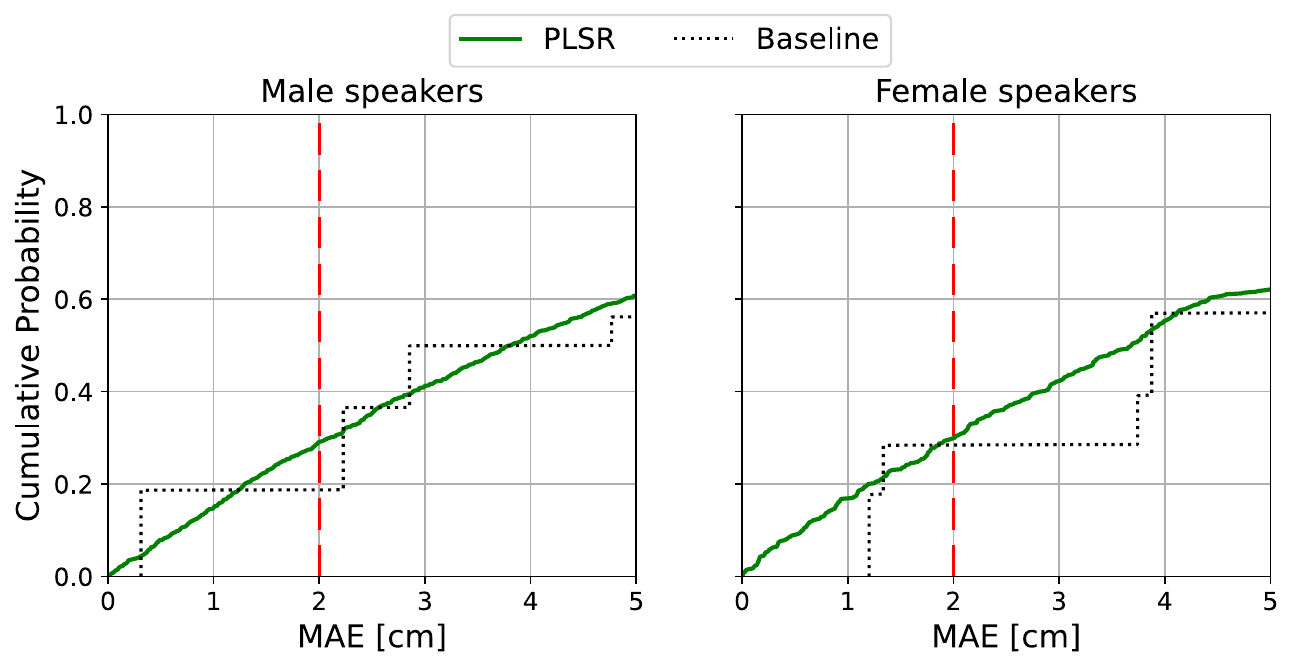}
  \caption{Empirical cumulative distribution function (eCDF) for height estimation error that falls within a predefined range.}
  \label{fig:ecdf}
\end{figure}

The aim of the first experiment is to analyze the benefits of using the presented HeightCeleb dataset. In this experiment, we assume that speaker gender is known. We believe this assumption holds for many real-life scenarios, where gender can be inferred from other modalities.
The results of this experiment are presented in Table \ref{tab:our_experiments}. %
At first, we compare two scenarios in which either the TIMIT (train) set or the HeightCeleb (train) set is used for training of both regression methods, while TIMIT (test) split is used for testing. Evidently, the MAE and RMSE results are significantly lower when HeightCeleb (train) is used for training instead of TIMIT (train), which shows the potential of the prepared HeightCeleb dataset for improving height predictions. Both MLR results are also notably better than for the Baseline algorithm, while PLSR proves to perform even better than MLR for the same training set. We also verified that the PLRS results of MAE are significantly different than those obtained by the Baseline method, as calculated with a paired t-test which resulted in $t(1119) =5.99$, $p=2.9 \times 10^{-9}$ for the male speakers and $t(559)=5.33$, $p=1.42 \times 10^{-7}$ for the female speakers, which allows to reject the hypothesis that the difference in MAE is not significant. Note that in case of the PLSR method, TIMIT (train) set is used as a validation set to fine-tune the hyper-parameter (number of components in the range 1 to 192) through minimizing MAE. Secondly, since we can use HeightCeleb (train) for training, we can additionally test the height prediction method on a full TIMIT dataset which includes both train and test splits, denoted in Table \ref{tab:our_experiments} as TIMIT (all). This way the statistical analysis of the results of the test is greatly improved. Furthermore, we also perform the test on the HeightCeleb (test) set as means of verifying performance on a large dataset. Finally note that for all experiments, as expected, the errors computed per speaker are in general lower than those obtained per utterance.
Table \ref{tab:results_others} presents the results of the second experiment, performed on the TIMIT (test) set, in which we compare three state-of-the-art methods \cite{height_gupta22_interspeech,height_singh2016short,height_rajaa2021learning} with the Baseline and our hierarchical method presented in Sec. \ref{ssec:method}, which at the first step performs gender classification with Logistic Regression trained on TIMIT (train), and in which PLSR is trained on HeightCeleb, with TIMIT (train) used for PLSR hyper-parameter tuning. These promising results show the usefulness of our HeightCeleb dataset, while we do not claim that our regression method is better than state-of-the-art approach from \cite{height_singh2016short} or \cite{height_rajaa2021learning}, as statistical tests are required to fully analyze the difference. In addition, taking into consideration problems with the reliability of height records in general, we believe that more in-depth analysis of errors should be performed and their significance studied.
Finally, we believe that the analysis of error distribution is more informative than comparing only their mean. Figure~\ref{fig:ecdf} depicts empirical continues probability distribution of an absolute error (for Baseline and PLSR model from Table \ref{tab:our_experiments}), which allows to assess the probability of getting the estimate within a predefined error threshold. It clearly shows, especially for the male speakers, that only a comparison of MAE or RMSE might be not enough to fully assess the usability of the trained model. For instance, the probability of making only a small error within a 2\,cm range (suggested as an acceptable error in \cite{brestoff2011challenging}) for PLSR increases from 19\% to 29\% for the male speakers and from 29\% to 30\% for the female speakers.
\begin{table}[t]
\centering
\caption{Comparison of speaker height estimation results from the literature (using single models) with the Baseline and our approach (with gender classifier) on the TIMIT (test) split.}
\label{tab:results_others}
\begin{adjustbox}{scale=0.95}
\resizebox{1\columnwidth}{!}{%
\begin{tabular}{ccccccc<{\hspace{3pt}}ccc<{\hspace{3pt}}ccc}
\toprule
\multicolumn{1}{c}{} & \multicolumn{1}{c}{} &\multicolumn{2}{c}{\textbf{MAE} {[}cm{]}} & \multicolumn{2}{c}{\textbf{RMSE} {[}cm{]}} \\
& \multicolumn{1}{c}{\multirow{-2}{*}{Method name}} & {Male} & {Female}  & {Male}    & {Female} \\ 
\midrule
\multicolumn{1}{c}{} & Baseline & 5.34 & 5.24 & 7.01 & 6.52 \\
\multicolumn{1}{c}{} & wav2vec 2.0 bi-encoder \cite{height_gupta22_interspeech} & 5.35 & 5.08 & 7.30 & 6.43 \\
\multicolumn{1}{c}{} & random forest regression\cite{height_singh2016short} & 5.0 & 5.0 & 6.7 & 6.1 \\
\multicolumn{1}{c}{} & wav2vec + LSTM \cite{height_rajaa2021learning} & 5.9 & 4.9 & 8.1 & \textbf{6.0} \\
\multicolumn{1}{c}{} & \textbf{Our (hierarchical)} & \textbf{4.81} & \textbf{4.73} & \textbf{6.21} & 6.18 \\
\bottomrule
\end{tabular}}
\end{adjustbox}
\end{table}

\section{Conclusions}%
This paper presents a new HeightCeleb dataset, an extension of the VoxCeleb dataset for speaker recognition with added information about the height of all 1251 speakers. 
We also show that a pre-trained neural speaker model that is followed by a regression method trained with HeightCeleb data achieves comparable results with state-of-the-art. %

\bibliographystyle{IEEEtran}


\end{document}